\documentclass[a4paper,11pt,oldfontcommands,oneside]{article}
\usepackage{latexsym}
\usepackage{amsmath}
\usepackage{amssymb}
\usepackage{graphicx}
\usepackage{enumerate}
\usepackage{natbib}
\usepackage{dsfont}
\usepackage{xcolor}
\usepackage[]{currvita}
\usepackage{comment}
\usepackage{geometry}
\usepackage{caption}
\usepackage{subcaption}
\usepackage{fancyhdr}
\usepackage{afterpage}
\usepackage{parskip}
\usepackage{authblk}
\usepackage[official]{eurosym}
\usepackage{pdfpages}
\usepackage{hyperref}
\usepackage{lineno}
\renewcommand{\sectionmark}[1]{\markright{\thesection.\ #1}}
\begin{document}


\renewcommand{\sectionmark}[1]{\markright{\thesection~~#1}}

\fancyhead[LE]{\thepage}
\fancyhead[RE]{\bfseries\sffamily\nouppercase  \leftmark}
\fancyhead[LO]{\bfseries\sffamily\nouppercase  \rightmark}
\fancyhead[RO]{\thepage}

\newgeometry{inner=3cm,outer=2.5cm,tmargin=3cm,bmargin=3cm}



\pagenumbering{gobble}

\newgeometry{inner=4.7cm,outer=2.5cm,tmargin=3.3cm,bmargin=5cm}

\pagestyle{fancy}
\fancyhead{}
\fancyhead[LE,RO]{\thepage}

\renewcommand{\sectionmark}[1]{\markright{\thesection~~#1}}
\renewcommand{\chaptermark}[1]{\markboth{\if@mainmatter\chaptername\ \thechapter~~\fi#1}{}}

\end{comment}

\fancyhead[LE]{\thepage}
\fancyhead[RE]{\bfseries\sffamily\nouppercase  \leftmark}
\fancyhead[LO]{\bfseries\sffamily\nouppercase  \rightmark}
\fancyhead[RO]{\thepage}

\newgeometry{inner=4.7cm,outer=2.5cm,tmargin=3.3cm,bmargin=3.3cm}

\title{Do Mature Economies Grow Exponentially?}
\author[1]{Steffen Lange\thanks{s.lange@knoe.org}}
\author[2]{Peter P\"utz\thanks{peter.puetz@uni-goettingen.de.}}
\author[3]{Thomas Kopp\thanks{thomas.kopp@agr.uni-goettingen.de (Corresponding author).}}
\affil[1]{Centre for Economic and Sociological Studies, Hamburg University; Konzeptwerk Neue \"Okonomie, Leipzig}
\affil[2]{Centre for Statistics, Georg August University G\"ottingen}
\affil[3]{Agricultural Market Analysis, Georg August University G\"ottingen}

\date{}
\maketitle

\begin{abstract}
Most models that try to explain economic growth indicate exponential growth paths. In recent years, however, a lively discussion has emerged considering the validity of this notion. In the empirical literature dealing with drivers of economic growth, the majority of articles is based upon an implicit assumption of exponential growth. Few scholarly articles have addressed this issue so far. In order to shed light on this issue, we estimate autoregressive integrated moving average time series models based on Gross Domestic Product Per Capita data for 18 mature economies from 1960 to 2013. We compare the adequacy of linear and exponential growth models and conduct several robustness checks. Our findings cast doubts on the widespread belief of exponential growth and suggest a deeper discussion on alternative economic grow theories.

\end{abstract}

\textbf{Keywords:} Gross Domestic Product Per Capita, exponential growth, linear growth, time series analysis.

\textbf{Acknowledgments:} 
\\Steffen Lange appreciates financial support from Hans B\"ockler Stiftung.
\\Thomas Kopp thanks Deutsche Forschungsgesellschaft (DFG) for funding through Collaborative Research Centre 990.


\clearpage

\begin{center}
\Large
\textbf{Do Mature Economies Grow Exponentially?}
\end{center}
\normalfont

\setcounter{page}{1}	

\section{Introduction}\label{chapone}
On November 2013, Lawrence Summer gave a seminal presentation at the annual IMF Research Conference on policy responses to the current economic crises. At this point, the great majority of economists studying the financial and the European economic crisis shared the view, that economic growth needed to be reestablished in order to solve the crisis. They ‘solely’ disagreed upon the way to get there, being divided into two camps – the austerity versus Keynesian stimulus proponents. Instead of arguing for either one of the two, Summers surprised his audience with an innovative perspective on the debate. He argued, that the US economy has entered secular stagnation - a new phase with constant low growth rates, and that this situation was about to stay.

Summers is not the first to observe the persistently low growth rates though. Several authors have, some already before the economic crises following 2007, argued that growth rates are falling in ‘mature economies’, that is, economies that typically were part of the first wave of industrialization  and have high per capita incomes today. While most economists reason that high growth rates can and should be regained \citep[p. 17]{reuter2000oekonomik}, according to this diverging analysis the low growth rates after the crisis are a continuation of decreasing growth rates for several decades. In fact, it is argued that post-war growth in mature economies depicts rather a linear than an exponential trend.

A significant number of authors argue that economic growth in early industrialized, high-income countries depicts linear instead of exponential growth \cite{Afheldt1994, Altvater2006, Bourcarde2006, Glotzl, Reuter2002, Wegweiser2014, HectorPollittAnthonyBarker2004}. These works are to a large degree grey literature and to our knowledge particularly strong within German-speaking debates. Even a subgroup of the ’German parliamentary commission on Growth, Welfare, Quality of Life’  expresses the view that macroeconomic growth in Germany has depicted linear instead of exponential growth in the past: ’Wie die Entwicklung des BIP in Deutschland in der Vergangenheit aber zeigt [...], hat das BIP dauerhaft lediglich mit konstanten jährlichen Zuwächsen zugenommen, also linear statt exponentiell’ (p. 136). To our knowledge only two studies investigated this question empirically for a set of countries. \cite{Bourcarde2006} find linear growth for 15 and exponential growth for 5 countries out of a sample of 20 countries. 
They look at macroeconomic growth instead of per capita growth though and do not apply rigorous statistical methods. \cite{Wibe2006} investigates the Gross Domestic Product Per Capita growth path of a total of 28 countries and country aggregates. Their results indicate a linear growth in the majority of the sample. We extend the work of \cite{Bourcarde2006} and \cite{Wibe2006} by using more recent data sets and more advanced statistical tools.

The question, whether economic growth is exponential, depicts decreasing growth rates or is of linear kind has important implications. First, it has serious implications for economic theory. The majority of economic (and in particular growth) theories explicitly or implicitly assume exponential growth rates, as we see below. If we observe linear growth, the theories and models need to be adjusted. Second, various policies depend on the size of growth:
\begin{description}
\item[$\bullet$ Employment:] If unemployment is the outcome of an interplay between productivity growth and economic growth \cite{Mankiw}, what labor policies are needed in case of linear economic growth?
\item[$\bullet$ Public debt:] If calculations of the GDP to debt ratio and of debt repayments are calculated based on predictions involving exponential growth rates, these become unreliable in case of linear growth \citep{Seidl2010a}.
\item[$\bullet$ Social security:] The financing of social security systems is usually shaped according to growth predictions based on exponential growth. In the situation of linear growth the common projections of the financing of security systems become unreliable and new strategies for long-term financing need to be developed \citep{Chancel2013}.
\item[$\bullet$ Environment:] There are two opposing views on the relationship between economic growth and environmental implications of economic activities. The first argues for a Kuznets-Curve like relation. In growing economies, environmental effects first increase and later decrease again \citep{Grossman1994}. This view is increasingly repudiated though \cite{Stern1996,Dinda2004,Kaika2013}. In the second view efficiency gains are not correlated with economic growth \citep{Stern2004}. Therefore, economic growth is associated with increasing environmental effects \citep{Victor2007,jackson2009prosperity,Kallis2012}. In particular in the case of climate change, growth is argued to increase environmental effects \cite{Kaika2013}. Linear growth changes the calculations on the relationship between economic growth and environmental impacts. This is likely to help explain why decoupling seems to have taken place in certain countries recently \citep{EuropeanEnvironmentAgencyEE} and it changes the necessary 
efficiency gains needed, which are usually calculated based on exponential growth \cite{jackson2009prosperity}.
\end{description}

In this article we first examine prominent orthodox and heterodox economic growth theories to what extend they do or do not predict exponential growth and why this is the case (section \ref{chaptwo}), leading to the development of an exponential and a linear growth function. The methods to test these functions empirically for 18 countries from 1960-2013 are described in section \ref{chapthree}. The corresponding results are presented in section \ref{chapfour}. Section \ref{chapfive} concludes.

\section{Literature}{\label{chaptwo}   
\subsection{Exponential growth as an inherent feature of modern growth theories}
Exponential economic growth is deeply routed within economic science and in particular within growth theories: ‘Mainstream macroeconomic theory is profoundly oriented towards an assumption of continuous, exponential GDP growth. Disruptions in economic activity, such as expansions and recessions, are perceived as deviations from the standard conception of a long-term stable macroeconomic growth path ’\cite[p. 1]{Seri2010}. The feature of exponential growth is part of growth theories throughout the history of economic thought. In the following, we give an overview of its role within several prominent schools of economic thought.
\subsubsection{Early modern growth theories}
After the classical contributions, which were very much engaged with growth theory (although they  termed it differently), economic growth became less prominent in economics. Following the Great Recession in 1930 the topic became central again. Harrod and Domar developed growth theories which entail concepts similar to Keynes' arguments at the time. Solow presented a first growth model in line with neoclassical thought and his model became central for neoclassical growth theory ever since.
\vspace*{6pt}
\\ \textbf{Harrod}
\\ \cite{harrod1939essay}  developed the concepts of warranted, actual and natural growth rates. The warranted growth rate ($g_w$) refers to the growth rate necessary for all savings to be invested. The growth rate depends on the savings rate ($s$) and the capital coefficient ($v$), as the later determines how much more can be produced per unit of investments: $g_w=\frac{s}{v}$ \footnote{The equations of this and the following sections are  based on the respective articles referred to.}. The actual growth rate depends on the actual investments taking place. Assuming that all savings are equally used for investments, the actual growth rate is given by 

\begin{equation}
g_a=\frac{s}{v_a}, 
\end{equation}

with the actual increase in the capital coefficient $v_a$. The natural growth rate can be regarded as an upper boundary of growth depending in particular on the labour supply due to population growth and the ‘work/leisure preference’ \cite[p. 30]{harrod1939essay}. The central concept is in his own words: ‘the geometric rate of growth of income or output in the system, the increment being expressed as a fraction of its existing level‘ \citep[p. 16]{harrod1939essay}. All three kinds of growth are expressed as growth \emph{rates}.

The explanation lies within the setup of the model. Harrod builds his argument on \emph{ratio} parameters, most importantly the savings rate and the capital coefficient. When savings are a percentage of production/income and investments need to equal savings, investments grow exponentially. The same holds for the capital coefficient: If the relation between capital and production increases by a certain percentage each year, the capital intensity of the economy grows exponentially. As a consequence, also production grows in an exponential manner.
\vspace*{6pt}
\\ \textbf{Domar}
\\ The work by \cite{domar1946capital} constitutes the second pillar of modern growth theory. In his approach, investments ($I$) on the one hand increase the capacity to produce ($dP$), dependent on the potential productivity of investments ($\sigma$) (Domar calls it the ‘potential social average investment productivity’ \cite[p. 140]{domar1946capital}). This is termed the \emph{capacity effect}: $\frac{dP}{dt}=\sigma I$. On the other hand, investments increase demand ($dD$). Its size depends on the size of investments and the savings rate, as a lower savings rate imply a higher multiplier effect. This so-called \emph{demand effect} is given by: $\frac{dD}{dt}=\frac{dI}{dt} \frac{1}{s}$. Setting the two equal gives the condition that the growth rate of investments ($g_I$) needs to be equal to the product of the savings rate and the potential productivity of investments: 

\begin{equation}
g_I=s \sigma.
\end{equation}

For Domar’s analysis the same holds as for Harrod’s: he builds his argument on ratio parameters as the potential productivity of capital and the savings rate.  Assuming a constant savings rate, exponential growth of investments is needed to make use of savings. If the productivity of investments stays constant, this implies also exponential economic growth.
\vspace*{6pt}
\\ \textbf{Solow model}
\\ \cite{solow1956contribution} developed, simultaneously to very similar work by \cite{Swan1956}, the first neoclassical growth model. In this theory, the growth rate of the capital stock ($\frac{\dot{k}}{k}$) is determined by investments, which depend on savings ($sF[k]$) and the depreciation of the capital stock ($\delta k$): $\frac{\dot{k}}{k}=s\frac{F[k]}{k}-\delta$. In his model, the capital per worker has decreasing marginal productivity. As the depreciation rate of the capital stock does not change, capital accumulation comes to an end. This is not the case if technological change is included. Therefore, Solow’s model is extended by introducing Harrod-neutral (i.e. labour-augmenting) technological change, which increases the productivity of labor. It implies a growth of the ‘effective labor’, which leads to a lower capital/labor ratio and subsequently increases the marginal productivity of capital. Hence, investments become profitable again. In this scenario, the steady state rate of economic growth 
is entirely determined by the speed of technological change $x$ \citep{Barro}: 

\begin{equation}
g=x.
\end{equation}

Accordingly, the model produces  exponential growth if a constant value for the speed of technological change is assumed.
\subsubsection{Neoclassical Growth Theories}
Since Solow's contributions, plenty of types and variations of neoclassical growth models have been developed. In the following, the three most prominent types are described. The neoclassical textbook model provides the microfoundations to Solow's model. The AK-model introduces constant returns to capital by including human capital, and models with imperfect competition endogenize technological change.
\vspace*{6pt}
\\ \textbf{Neoclassical Textbook Growth Model}
\\While Solow assumed a certain savings rate and a certain investment behavior of firms, neoclassical growth models are based on the behavior of a representative household and a representative firm. Households maximize utility due to a utility function. Their savings depend on their preferences and the interest rate. Firms maximize profits. They invest until the marginal productivity of capital is equal to the real interest rate plus the rate of depreciation. Subsequently, in these models savings and investments are brought into equilibrium via the interest rate \citep[Chapter 2]{Barro}.

The equilibrium growth rate of the capital stock is similar to the Solow-model. The change in the capital stock is determined by the difference between output and consumption. Additionally, the capital needed for depreciation and due to technological change is subtracted: $\frac{\dot{\hat k}}k=\frac{f(\hat k)-\hat c}{\hat k} - (\delta+x)$.

The growth of per capita income ($g$) depends - as in the Solow model – primarily on the rate of technological progress ($x$) though: ‘the steady-state per capita growth rate equals the rate of technological progress, $x$, which is assumed to be exogenous’ \cite[ p. 205]{Barro}. Also, ‘A greater willingness to save or an improvement in the level of technology shows up in the long run as higher levels of capital and output per effective worker but in no change in the per capita growth rate’ (p. 210):

\begin{equation}
g=x.
\end{equation}

The neoclassical textbook model therefore leads to the same conclusion concerning exponential growth as did the Solow-model: As long as technological change is assumed to be of a constant \emph{rate}, economic growth is exponential.
\vspace*{6pt}
\\ \textbf{Endogenous growth I: The AK-model}
\\ The unsatisfactory result from neoclassical growth theories, that economic growth depends purely on an exogenously given rate of technological progress in the neoclassical growth theories motivated the development of models which explain continuous growth differently. In the AK-model capital has constant instead of diminishing marginal returns, due to the fact that capital is understood in broader terms, including human capital. Production in this model depends on the technological state ($a$) and the amount of capital ($k$): $y=ak$. Growth of capital per capita ($\frac{\dot{k}}{k}$) and income per capita ($\frac{\dot{y}}{y}$ or $g$) depend, without technological change, on the size of net investments which depend on the savings rate and the depreciation rate \citep[Chapter 1 and 4]{Barro}:

\begin{equation}
g=\frac{\dot{y}}{y}=\frac{\dot{k}}{k} = s A - \delta
\end{equation}

The AK-model is therefore also designed to depict exponential growth. The reason lies within the ratio of the savings rate that determines capital accumulation with constant returns. Thereby investments grow exponentially and so does per capita income.
\vspace*{6pt}
\\ \textbf{Endogenous growth II: Imperfect competition}
\\Endogenous growth models with imperfect competition assume markets (of monopolistic competition) \citep{Aghion1998} in which firms invest into new production methods or the improvements of the existing methods because they have a temporary patent on the new method which allows them to make profits. 
The rate of technological change and hence the rate of economic growth depends on how fast new intermediate goods are invented. This is determined by several factors. First, the preferences of households concerning consumption and savings ($\theta$ represents the households’ willingnesses to have different consumption levels over time and $\rho$ stands for the time preference of the households) influence the amount of resources put into developing new technologies. A second set of parameters influence the speed of technological change: The price of inventing new technologies ($\eta$) and the size of the mark-up ($\alpha$) a firm can put on the new technology. Finally, the state of technology ($A$) and the amount of the labor employed ($L$) influence the growth rate. The following equation gives an example of the determinants of economic growth \citep[Chapter 6 and 7]{Barro}:

\begin{equation}
\frac{\dot{Y}}{Y}=\frac{1}{\theta} [\frac{L}{\eta} A^{\frac{1}{1-\alpha}} (\frac{1-\alpha}{\alpha}) \alpha^{\frac{2}{1-\alpha}} - \rho].
\end{equation}

Here, growth is also represented as a rate (this time it is not per capita growth but macroeconomic growth). A change in each of the factors mentioned therefore alters the rate of growth. Again, constant parameters lead to exponential growth in this model. 
\subsubsection{Heterodox growth theories}
In neoclassical frameworks growth depends on two central aspects: savings lead to investments and technological change increases labor productivity. Heterodox growth theories also regard these factors as central, but analyze the underlying mechanisms very differently. Generally speaking, investments are not only determined by savings but also other factors and technological change either depends on investments or is the effect of market competition.
\vspace*{6pt}
\\ \textbf{Marxian}
\\In Marx's theory, firms buy labor (variable capital), materials and physical capital (constant capital) in order to manufacture products that they sell on a competitive market. Firms can make profits due to the exploitation of labor, i.e. because workers are paid due to their cost of reproduction that usually is below the value of their labor power \citep{Harvey2006}. As firms sell their products on a competitive market they are forced to invest their profits into new production technologies. In case they do not so, they are outcompeted by their rivals who did and therefore are able to offer products at a lower price \citep{mandel1969marxist}. The degree of competition is the prime determinant of the share of profits ($a$) reinvested ($a=\frac{I}{\pi}$). The profit rate ($r$) is primarily due to the bargaining process between capital and labor ($r=\frac{\pi}{K}$). The growth rate of the capital stock, which is equivalent to the growth rate of output ($g$) depends on these two factors \citep{Hein2002}:
\begin{equation}
g=\frac{I}{K}=ar.
\end{equation}

Investments go hand in hand with technological change. The introduction of new technologies changes the organic composition of capital. It increases the ratio between constant and variable capital. Investments hence have ambiguous effects on employment. On the one hand they increase the demand for labor, as production increases. On the other hand, the labor-coefficient decreases, which decreases the demand for labor. Marx argues, that the second mechanism guarantees a continuous existence of a 'reserve army', i.e. a continuous availability of workers.

Whether the Marxian theory suggests an exponential growth pattern or not is up to constant debate. It is disputed under the term “Tendency of the rate of profit to fall”. On the one hand it is argued, that the change in the organic composition of capital decreases the profit rate. The intuitive reason is, that surplus is generated due to the exploitation of labor power. Therefore, assuming everything else constant, using relatively less labor in the production process also implies less possibility to generate surplus and profits. On the other hand, if the surplus rate (the ratio between surplus and variable capital) increases, this tendency can be countervailed \citep{Sweezy1943a}. Hence, the Marxian analysis leaves it open, whether economic growth takes place at a certain rate (and hence is exonential) or whether the growth rate declines.
\vspace*{6pt}
\\ \textbf{Kalecki}
\\ \cite{Kalecki1987} developed a growth theory that builds on both Marxian and Keynesian analyses and concepts. Investments are at the center of his argument. The economy is divided into three sectors which produce (1) capital goods for investments ($I$), (2) consumption goods for capitalists ($CK$) and (3) consumption goods for workers ($CW$). The size of the economy $Y$ is therefore given by $Y=I+CK+CW$. In each sector capitalists earn profits ($P$) and workers earn wages ($W$). There are no savings out of wages. Investments and the consumption of capitalists determine profits: $P=I+CK$. Investments determine growth. Its effect depends on the wage share ($w$) and the consumption rate of profits ($q$):$\Delta Y=\frac{\Delta I}{(1-w)(1-q)}$.

According to \cite{Kalecki1987} there are five central determinants of investments. (1) Firm’s savings ($S$) induce entrepreneurs to invest more, as they have more financial means. (2) An increase (decrease) in firm’s profits $\frac{\Delta P}{\Delta t}$ stimulate (attenuate) them to increase investments because production has become more profitable. (3) A change in the capital stock $\frac{\Delta K}{\Delta t}$ increases/decreases investments because additional profits due to investments are lower/higher for a higher/lower level of existing capital. (4) An increase in production in the past $\frac{\Delta Y}{\Delta t}$ induces investments because inventories are proportionate to production. (5) Finally, investments may change due to various long run changes ($d$) in the economy (technological change, interest rate, company share earnings): $I_{t+1}=aS_t + b\frac{\Delta P}{\Delta t} - c\frac{\Delta K}{\Delta t} + e\frac{\Delta Y}{\Delta t} + d$.

Kalecki further argues, that the change in profits, the change of the capital stock and the change in production all primarily depend on the investments of the past. Subsequently, investments are a function of past investments, several behavioral parameters (covered by $a, b, c, e, q, p$) and the long-term changes ($d$): $I_{t+1}=aI_t - c\frac{\Delta K}{\Delta t} + \frac{1}{1-q} (b + \frac{e}{1-w}) \frac{\Delta I}{\Delta t} + d$

Investments are the sum of a fraction of the past investment level ($aI_t$), the change of investments in the past multiplied by some constant ($\frac{1}{1-q} (b + \frac{e}{1-w}) \frac{\Delta I}{\Delta t}$) and the change of the capital stock multiplied by some constant ($c\frac{\Delta K}{\Delta t}$). Assuming constant behavioral parameters, economic growth in Kalecki’s theory therefore tends to also be exponential. This can best be illustrated by assuming that $a=1$ and abstracting from depreciation of the capital stock. Defining $f=\frac{1}{1-q} (b + \frac{e}{1-w})$, we get 

\begin{equation}
I_{t+1}=I_t + (f-c) \frac{\Delta I}{\Delta t}.
\end{equation}

As long as $f>c$, the growth of investments is therefore of exponential nature. Assuming a constant capital coefficient and constant population, economic growth is therefore also exponential. The result can only be different, if the parameters change over time.
\vspace*{6pt}
\\ \textbf{The standard Keynesian growth model}
\\ \cite[chapter 7]{hein2004verteilung} develops a standard Keynesian model of growth and distribution. In this model there are only savings out of profits, so the savings rate (savings/capital stock, $s$) is equal to the savings ratio out of profits ($s_\pi$) multiplied by the profit rate (profits/capital stock, $r$): $s= s_\pi r$. The growth rate of the capital stock is firstly determined by animal spirits ($\alpha$) (e.g. \citep{Fontana2015}) and, secondly, by the investment reaction ($\beta$) to the profit rate ($r$): $g=\alpha + \beta r$ (e.g. \cite{Kalecki1987}). A steady-state growth rate exists, where the growth rate ($g$) is equal to the savings rate: $g=s$. Combining these equations one gets the equilibrium growth rate
\begin{equation}
g=s= \frac{s_\pi \alpha}{s_\pi - \beta}
\end{equation}

In this case, the growth rate stays constant, if the other parameters also stay constant. Again the model depicts exponential growth.
\subsection{Explanations for diminishing growth rates}
Discussions on an end to economic growth have existed at least since the beginning of modern economic theory. All the major classical economists had a concept of the steady-state, that marked an end to economic expansion \citep{Luks2001}. The term secular stagnation was coined by \citep{Hansen1939} in the Great Depression. He argued that a limit to geographical expansion, seizing population growth and less capital-intensive technologies resulted in lower investment rates and lower growth. In the first controversy on this issue, \cite{Schumpeter1939} on the other hand argued that unfavorable business conditions for entrepreneurs were the reason for lacking growth rates. While this first discussion was ended by World War Two and the economic expansion of Western Europe and Northern America afterwards, it came up again with the stagflation in the 1970s. \cite{Sweezy1982} argued, that again the lack of investments was the reason that, while recovery of a growth path was possible in theory, “nothing like that is 
visible on the horizon now” \citep[p. 9]{Sweezy1982}. The discussion has been taken up again after the global financial crisis after 2007 and has gained momentum after the speech by Summers mentioned in the beginning of this article. However, there are also arguments for declining and/or linear growth which are not associated to the term secular stagnation. In the following several different reasonings, which mostly also refer to one or several of the theoretical frameworks developed in section \ref{chaptwo}, are discussed.
\subsubsection{Slower technological change}
As we have seen above, technology plays a central role in explaining economic growth. Hence, it is not surprising that some authors see slower technological change as being the reason for  declining growth rates. It is argued, that the technological progress in recent decades has increased labor productivity less than in the decades before – and that it is the nature of the technological change that is the explanatory of less growth. \cite{Gordon2012} argues that capitalism has experienced three industrial revolutions (IR): ''The first (IR \#1) with its main inventions between 1750 and 1830 created steam engines, cotton spinning, and railroads. The second (IR \#2) was the most important, with its three central inventions of electricity, the internal combustion engine, and running water with indoor plumbing, in the relatively short interval of 1870 to 1900. [...] The computer and Internet revolution (IR \#3) began around 1960 and reached its climax in the dot.com era of the late 1990s`` \citep[p. 1-2]{
Gordon2012}.
For him the divergent impacts of the innovations on labor productivity explain different growth rates. Much of the high growth rates in the 1950s to 1970s can be explained by a combination of applications of technologies from the second and third industrial revolution. Since 1972 the effects of the implication of technologies from the second industrial revolution have faded out and the remaining effects of the third revolution are a major explanation for the low growth rates that we observe today in mature economies\citep{Gordon2014}.
This analysis directly refers to the early growth theories with exogenous technological growth. While Gordon takes into account that the speed of technological change is also influenced by other factors, he argues for a baseline rate of technological change \cite{Gordon2014a}. This baseline is exogenously given.
\subsubsection{Labor and human capital}
Within the recent discussion on secular stagnation there are two arguments concerning labor. First, the average number of working hours per capita has decreased and is predicted to continue to do so in the future. While the average working hours per worker have declined in Europe (not so much in Northern America), the participation rate of women has increased \cite{Maddison2006}. As someone also has to do the reproductive work, this leaves little room for an extension of labor hours in the future \citep{Adam2013}. The major reason for declining average working hours is the demographic change though, leading to an increase in the dependency ratio. This is anticipated to continue in the future for almost all countries investigated in this paper. Based on these findings, \cite{Johansson2012} come to the conclusion that 'Population ageing, due to the decline in fertility rates and generalized gains in longevity, has a potentially negative effect on trend growth as it leads to a declining share of the working age 
population as currently defined (15-64 years)` \citep[p. 13]{Johansson2012}.
Second, it is argued that the advantages from education to increase the productivity of workers have declined over the past decades and that this trend is likely to continue. \cite{Gordon2014a} see the major reasons in certain institutions. They argue, that future “increases in high school completion rates are prevented by dropping out, especially of minority students” (p. 51) and that the inability for many to finance academic studies is a major problem to further improve education. Similarly, \cite{Eichengreen2014} argues that the US government has spend too little on public education over the past decades. 
The second argument relates to the views of endogenous growth theories including the role of human capital. From this perspective, the investments into human capital have been low, so that growth is also low. Another explanation could be, that marginal returns to human capital are diminishing. To our knowledge, this argument has not been made yet though.
\subsubsection{Insufficient investments}
Investments play a crucial role for economic growth, as was the case in all growth theories covered above. They are important for capital deepening and the application of new technologies. Within the recent literature on secular stagnation, several reasons for insufficient investments have been pointed at: (1) \cite{Eichengreen2014} argues that low investments by the government, in particular in infrastructure, have been a major reason for overall low investments in the USA. (2) The decrease in population growth decreases overall investments and therefore slows down the application of new technologies – hence dampens technological change \citep{Hansen1939,Krugman2014}. (3) It is argued that working aged people buy relatively more capital-intensive goods (in particular housing) and that the demographic change has therefore decreased the demand for such goods, which also dampens overall investments \citep{Hansen1939,Krugman2014}. The causes of low investments are hence argued to be found in low demand or low 
governmental investments. Accordingly, these views refer to the heterodox and in particular the Keynesian theories on growth.
\subsubsection{Non-competitive market structures}
Another explanation for decreasing investment rates refers to market structures. This work builds on marxian reasoning as outlined above. While Marx assumed perfect competition in that argument, \cite{Baran1966a} argue that post-war capitalism was marked by increasing concentration of market power. Therefore oligopolistic and monopolistic market structures had become the dominant form of market structures. \cite{Steindl1976} argues that such structures introduce several mechanisms which overall lower investment and growth rates. The mono-/oligopolistic structures increase profit rates within the concentrated sectors and thereby lower aggregate demand (assuming higher savings out of profits than out of wages). At the same time, investments in these sectors are as monopolists maximize profits with investments and production below the level within competitive markets. Due to the low demand, other profitable investment opportunities are scarce though, leading to an overall reduced level of investments and an 
equilibrium below potential output. \cite{Foster2014a} comes to the conclusion that the outcome of concentrated markets is 'a chronic condition of secular stagnation' \citet[p. 87]{Foster2014a}.
\subsubsection{Consumption demand}
Consumption is the second major component of aggregate demand. Next to investments, a slow increase in consumption can therefore be another major cause for low growth. There are again two central arguments. 
Within the discourse on secular stagnation, it is argued that consumption slackens due to high income and wealth inequalities and the accompanying household debts. \cite{Summers2014} point out that “changes in the distribution of income, both between labor income and capital income and between those with more wealth and those with less, have operated to raise the propensity to save, as have increases in corporate-retained earnings” (p. 69). Summers argues that this is a reason for the decline of the interest rate. It also decreases consumption though: “Rising inequality operates to raise the share of income going to those with a lower propensity to spend” \cite[p. 33]{Summers2014a}. \citet{Krugman2011} argues on the other hand, that at least before the crisis, the reason for low growth was not high savings (as the savings rate was not high) but the trade-deficit in the USA . Nevertheless, Summers' argument – that is also being made by other such as \cite{Gordon2014a} and \cite{Eggertsson2014} can be 
maintained when looking at other countries of our sample.
The second argument refers not to a lower propensity to consume for higher income-earners at a certain point in time but over time. The argument is, that due to several mechanisms, with increasing average income, people consume a lower share of their income. This leads to only slowly increasing aggregate demand and subsequently to low growth. \cite{reuter2000oekonomik} made this argument long before the global financial crisis. He argues that the satisfaction of needs combined with institutional limits to exponentially increasing consumption are responsible for only slowly increasing consumption and slow growth. In fact, this line of argument goes back to Keynes' work. \cite{keynes1963economic} argued that with increasing average income, people would choose leisure time over further consumption increases. These lines of argument therefore directly correspond to the Keynesian growth theories discussed above.
\subsubsection{Slower increase in energy use}
In most economic growth theories, the environment is either seen as a source for natural resources and/or a sink for emissions. Accordingly, the use of natural resources is mostly modeled as a third input into the production function. In neoclassical theory and environmental economics, natural resources are assumed to be substitutable by physical capital. In ecological economics on the other hand, substitution between them is regarded to be very limited as they are primarily complements. 
Based on this analysis, several authors have argued that by accounting for exergy and/or useful work, economic growth can be accounted for to a much higher degree than by traditional procedures. \cite{Ayres2005a} find that useful work can explain much of past economic growth. Interestingly, they can explain the growth until the mid-1970s almost entirely, while afterwards, this is less the case. From this point in history, the increase of useful work slows down significantly. Economic growth slows down less so. An interpretation could be, that the increase in energy prices has led to a new technological trajectory, which is depicted by less use of energy and at the same time less productivity growth. At the same time, there are many other changes going on and this phenomenon needs further investigation.

\section{Methods and Data}\label{chapthree}


As we have seen, most prominent growth theories argue for an exponential growth pattern. This is represented by one or several ratios, be it the savings ratio, the speed of technological change, the profit rate or the potential productivity of investments. Accordingly, a Gross Domestic Product Per Capita (GDPPC) series underlying the growth models covered so far (excluding the Marxian theory, which is not clear in this respect) can be represented within a simple regression framework by the following term:

\begin{equation}
\label{equation_expt}
GDPPC_t = b_0   b_1^t+\varepsilon_t = b_0   \left(1+r \right)^t+\varepsilon_t, \; t=0, \dots, T-1,
\end{equation}
where $b_0$ denotes the starting value in $t=0$ of the series with $T$ observations, $b1$ (and thus $r$) determines the growth of the series, and $\varepsilon_t$ is the error term for the observation in $t$.

As an alternative hypothesis we test the exponential model against the simplest model of diminishing growth rates, which is a model of linear growth:

\begin{equation}
\label{equation_lin}
GDPPC_t= b_0 + b_1t + \varepsilon_t=b_0 + \left(1+r \right)t + \varepsilon_t, \; t=0, \dots, T-1,
\end{equation}
where $b1$ (and thus $r$) now enters the equation in a linear fashion which corresponds to a constant increase of the GDPPC .

In order to examine these growth theories on an empirical basis, we look at the economic development for a set of 18 mature economies from 1960 to 2013. In particular, we decided for investigating yearly real GDPPC series for the group of Western and Southern European countries and Western offshoots (as defined by \citet[Appendix B]{Maddison2006}). Germany was left out of the sample in order to avoid problems of aggregation during the period before re-unification in 1990. Luxembourg was added, which Maddison includes in the group of “Small West European Countries” \citep[p. 179]{Maddison2006}. A full overview of the selected countries can be found below in table \ref{table1}. The rationale behind the starting point in 1960 is that effects of World War Two should have vanished by then. As \citet{Crafts1996a} argue, 'In five years at most, Europe recovered the ground lost relative to the highest prewar income levels. It is, thus, quite safe to place the end of the first phase of reconstruction and the beginning of a new era in the history of European economic growth in 1950.' \citep[p. 3]{Crafts1996a}. Since the data is only available since 1960, we start from there.

It has been argued that the  second major reason for high growth rates in Europe (including the majority of countries of our sample) after World War Two was convergence. The US-American economy had introduced new production methods characterised by higher labour productivities over the previous decades, which European countries had not. From the end of World War Two until the end of the 1960s the introduction of such technologies in European countries facilitated the high growth rates \citep{Eichengreen2008}. In order to exclude this effect, we additionally execute the empirical investigation for the period 1970-2013.

Due to partly differing GDPPC measures, we analyse two data sets: World Bank GDPPC series in US\$ with 2005 prices and Conference Board GDPPC series in US\$ with 2014 prices. Since the results are very similar, we only report the results for the first data set mentioned above.\footnote{The data sets, non-reported results and the software code can be obtained from the authors upon request.} In this data set, GDPPC data for New Zealand and Switzerland are only available from 1970-2013.

In a preliminary analysis we compare for each country the coefficients of determination $R^2$ between a regression of the GDPPC series on a linear and on an exponential time trend.\footnote{Note that this approach is essentially equivalent to the comparison of the log-likelihood between a non-transformed and a log-transformed GDPPC series as done by \citet*{Wibe2006} who used GDPPC data up to 2005.} The first regression corresponds to equation (\ref{equation_lin}) and the latter to equation (\ref{equation_expt}). In order to obtain the optimal exponential growth rate within the regression framework, the exponential growth model is estimated for a grid over 50 equidistant values between 0 and 0.06 for $\widetilde{r}$\footnote{Clearly, all countries in our sample exhibit a positive growth within the time frame under consideration.}  and the highest $R^2$ is selected.

This procedure represents a rather descriptive approach of comparing linear and exponential growth of the GDPPC . Making credible quantitative statements about the data generating process behind the time series is, however, only valid for stationary series. Only after taking account for possible unit roots and autocorrelation of time series it is reasonable to compare the adequacy (in terms of predictive ability) between different time series models. Note that these issues were not tackled by \citet*{Wibe2006} and \citet*{Bourcarde2006}. We therefore apply the Box-Jenkins method \citep{Box2011} to find suitable models for the series at hand. The first step is to determine the order of integration of the time series. Initialised by \citet*{Nelson1982}, there has been an extensive debate on whether macroeconomic time series are trend-stationary or follow a unit root process with a potential drift, see for instance \citet*{Perron1989}, \citet{Kwiatkowski1992} and \citet*{Cuestas2011}. The latter view seems to be the more prominent in the literature as most authors apply unit root and co-integration techniques to macroeconomic time series like the GDPPC. In order to check whether unit root methods are also required for the data set at hand, we conduct for each country the augmented Dickey-Fuller test for both the original time series and the logarithm (log) of GDPPC series after removing their linear time trends. Note that removing a linear time trend from a log-transformed time series corresponds to the deletion of an exponential time trend of the original series. A rejection of the null hypothesis suggests the absence of a unit root and therefore trend-stationarity. Likewise, we conduct the KPSS test for the same series, yet the null hypothesis for this test is trend-stationarity. The results of these tests can be found in Table \ref{table_appendix} in the \hyperref[chapsix]{appendix}. As we find no strong evidence against a unit root and for trend-stationarity in both the original and the log transformed GDPPC in any of the countries except for Switzerland,\footnote{We chose a significance level of 5\%.} we generally assume the series to follow unit root processes with drift terms. We are aware of the heterogeneity of the countries and the weaknesses of the underlying tests, as pointed out in \citet*{Cuestas2011}. In particular, our simple time trend models might not capture true nonlinear patterns, for instance caused by structural breaks. Nevertheless, as our test results are quite unambiguous, we are confident that the aggregate findings of the following analysis should be credible.

Assuming a unit root, the next step is to determine the orders $p$ and $q$ of the autoregressive process and the moving average process in a suitable ARIMA ($p,1,q$) model with drift. In order to do so, we use the \textit{auto.arima} function of the \textit{R} package \textit{forecast} \citep*{Hyndman2008a}. More specific, we choose among all candidate models with maximum lag orders $p=q=3$ the most appropriate one with respect to the Akaike Information Criterion for finite sample sizes (AICc).\footnote{We also use the Bayesian Information Criterion (BIC) for model selection in this step and obtained very similar results. For a comprehensive discussion on model selection criteria in time series models, see \citet[chap. 7]{Hyndman2008b}.} This procedure achieves an appropriate compromise between a lack-of-fit to the data and a too complex model. As we are estimating an ARIMA($p,1,q$) with drift, we actually model the first differences of GDPPC series. Allowing for a potential exponential growth, this leads in the simplest case with $p=q=0$ to the model

\begin{equation}
\Delta GDPPC_t=GDPPC_t-GDPPC_{t-1}=\widetilde{b}_0 \left(1+\widetilde{r} \right) ^{t-1}+\widetilde{\varepsilon}_t, \; t=1, \dots, T-1
\end{equation}

where the errors $\widetilde{\varepsilon_t}$ are assumed to be independent and identically distributed and $\widetilde{b_0}$ describes the difference between the first two observations which grows over time at the rate $\widetilde{r}$. A constant and thus linear growth is given for $\widetilde{r}=0$. We exploit exactly this property to decide for either a linear or an exponential growth model by using a similar approach as above: first, for each of the 50 equidistant values between 0 and 0.06 for $\widetilde{r}$, the most suitable ARIMA ($p,1,q$) model with drift with respect to the AICc is chosen. In a second step we decide for the final model with a corresponding growth rate (which we subsequently refer to as optimal model with optimal growth rate) by comparing the value of several selection criteria amongst all 50 ARIMA ($p,1,q$) models chosen in the first step. The predictive ability of different time series models is often compared by the accuracy of pseudo-out-of-sample-forecasts. In those cases, the particular models are fitted to a subsample of the data (e.g. all observations except for the last 5 years) and the last data points are predicted based on the estimated model parameters. This procedure essentially tells us which model forecasts better at the end of the sample. However, we are aware of a potentially big influence of the recent financial crisis on the results if we solely compared forecast accuracy for the last years. Thus, for our second step of final model selection, we use the Akaike Information Criterion (AIC), which includes all observed data points and can thus be seen as a measure for predictive ability for the whole sample \citep*[chap. 7]{Hyndman2008b}. Furthermore, we also base our selection of the optimal model on the AICc and the more parsimonious Bayesian Information Criterion (BIC). Despite the advantages of such model selection criteria, the results are still likely to depend on the chosen sample. For this reason we repeat the analysis with excluding the recent financial crisis, i.e. considering only the years 1960-2007. As a further robustness check, we discard the first 10 years of the sampling period and thereby restrict our analysis to the years 1970-2013, in order to exclude the effect of convergence, as argued above.

\section{Results}\label{chapfour}



The results of the $R^2$ comparison between the best exponential fit (chosen among all 50 grid values for the growth rate $\widetilde{r}$) and the linear fit can be found in table \ref{table1}. For the majority of countries, the linear model fits the data more accurately than the exponential one. In a couple of countries, the evidence points into the opposite direction, with the differences between the models being rather small in these cases, however.

\begin{table}[t b]
\begin{center}
\begin{tabular}{lccc}

\hline
Country & $R^2$ exp. fit & $R^2$ lin. fit & diff. \\ 
\hline
Australia & 0.9910 & 0.9755 & -0.0155 \\ 
Austria & 0.9709 & 0.9951 & 0.0241 \\ 
Belgium & 0.9544 & 0.9886 & 0.0341 \\ 
Canada & 0.9597 & 0.9819 & 0.0222 \\ 
Denmark & 0.9491 & 0.9774 & 0.0283 \\ 
Finland & 0.9578 & 0.9728 & 0.0150 \\ 
France & 0.9329 & 0.9779 & 0.0449 \\ 
Italy & 0.8906 & 0.9526 & 0.0620 \\ 
Luxembourg & 0.9509 & 0.9356 & -0.0153 \\ 
Netherlands & 0.9686 & 0.9824 & 0.0138 \\ 
New Zealand & 0.9452 & 0.9246 & -0.0206 \\ 
Norway & 0.9526 & 0.9836 & 0.0310 \\ 
Portugal & 0.9198 & 0.9711 & 0.0513 \\ 
Spain & 0.9403 & 0.9699 & 0.0296 \\ 
Sweden & 0.9774 & 0.9711 & -0.0062 \\ 
Switzerland & 0.9566 & 0.9537 & -0.0029 \\ 
United Kingdom & 0.9702 & 0.9601 & -0.0101 \\ 
United States & 0.9760 & 0.9879 & 0.0119 \\ 
   \hline

\end{tabular}
\end{center}
\protect\caption{$R^2$ for exponential and linear model and difference between them. A positive difference indicates the preference of the linear fit. Time horizon: 1960-2013.}
\label{table1}
\end{table}

Table \ref{table2} depicts the optimal growth rate of the optimal ARIMA ($p,1,q$) model (with respect to the model selection criteria AIC, AICc and BIC after our two-step selection procedure) explaining the observed GDPPC series for each country. 

\begin{table}[t b]
\begin{center}
\begin{tabular}{lccc}
\hline
Country & AIC & AICc & BIC \\ 
\hline
Australia & 0.0110 & 0.0110 & 0.0110 \\ 
Austria & 0.0000 & 0.0000 & 0.0025 \\ 
Belgium & 0.0000 & 0.0000 & 0.0000 \\ 
Canada & 0.0000 & 0.0000 & 0.0000 \\ 
Denmark & 0.0000 & 0.0000 & 0.0000 \\ 
Finland & 0.0000 & 0.0000 & 0.0000 \\ 
France & 0.0000 & 0.0000 & 0.0000 \\ 
Italy & 0.0600 & 0.0600 & 0.0000 \\ 
Luxembourg & 0.0037 & 0.0037 & 0.0037 \\ 
Netherlands & 0.0000 & 0.0000 & 0.0000 \\ 
New Zealand & 0.0135 & 0.0135 & 0.0135 \\ 
Norway & 0.0000 & 0.0000 & 0.0000 \\ 
Portugal & 0.0000 & 0.0000 & 0.0000 \\ 
Spain & 0.0000 & 0.0000 & 0.0000 \\ 
Sweden & 0.0025 & 0.0025 & 0.0025 \\ 
Switzerland & 0.0012 & 0.0012 & 0.0012 \\ 
United Kingdom & 0.0025 & 0.0025 & 0.0025 \\ 
United States & 0.0000 & 0.0000 & 0.0000 \\ 
   \hline

\end{tabular}
\end{center}
\protect\caption{Optimal growth rates leading to the optimal ARIMA ($p,1,q$) model with drift with respect to different model selection criteria. A growth rate of (close to) zero implies a linear growth model. Time horizon: 1960-2013.}
\label{table2}
\end{table}

The linear growth model with $\widetilde{r}=0$ is preferred for 11 out of 18 countries with respect to the AIC. Note that some optimal growth rates are not equal but close to zero and thus far from the magnitude that is usually talked about when discussing growth rates, indicating a quasi-linear growth. Such an example is Sweden which exhibits an optimal growth rate of only 0.25\%. The behaviour of the AIC over all grid values for the growth rate $\widetilde{r}$ is shown in Figure \ref{figure1} for the case of Sweden. It can be seen that the lowest AIC at the growth rate 0.25\% is very close to the AIC for the linear model with $\widetilde{r}=0$, whereas it becomes considerably larger for increasing values of $\widetilde{r}$.\footnote{Naturally, a quantitative comparison between a linear and an exponential growth model for a single country is in general not possible if the linear model is chosen, since there is no comparable best exponential growth model.}

\begin{figure}[t b]
\centering
\includegraphics[width=140mm]{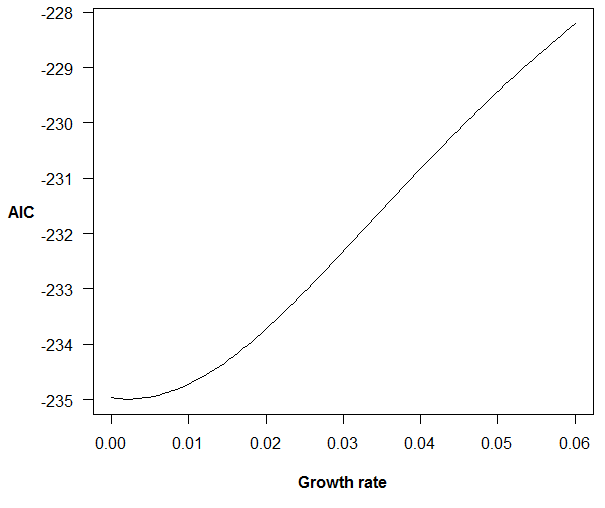}
\protect\caption{AIC values for the ARIMA ($p,1,q$) models chosen for different values of the growth rate $\widetilde{r}$, illustrated for Sweden. A lower AIC indicates a better predictive ability.} 
\label{figure1}
\end{figure}

The results are fairly robust for the chosen model selection criteria, with the only exception being Italy: here, the AIC and AICc indicate an exponential growth model with the largest growth rate $\widetilde{r}=0.06$ on our grid, whereas the BIC prefers the linear model. A closer look at the optimal ARIMA($1,1,3$) model chosen by AIC and AICc shows a near unit root autoregressive coefficient of 0.9989 and an implausible negative coefficient for $\widetilde{ b_0}$ which would imply negative exponential growth.\footnote{For all other countries, the estimate for $\widetilde{b}_0$ is positive and statistically significant at the 5\%-level. The results are available upon request.} In contrast, the BIC points towards an ARIMA ($0,1,1$) model with a positive linear trend.  These unstable results suggest that the development of Italy’s GDPPC is hardly predictable, at least within our model class. 

 The financial crisis following the year 2008 led to a deep recession of the world economy and was an extraordinary economic event. 
Since, in our case, the model selection criteria are likelihood-based measures relying on the normal distribution, outliers strongly affect the choice of the optimal model and the corresponding optimal growth rate for differenced series like the ARIMA($p,1,q$) processes.\footnote{More specific, the squares of deviations from the mean enter into the estimation. Thus, an outlier in a differenced series, which has little variation otherwise, has heavy weight on the obtained results. For a better technical understanding of the maximum likelihood estimation procedure, the reader is referred to background literature, for instance \citet{Fahrmeir2013}.} We therefore execute the analysis described above once again, excluding the respective years of the crisis.  As table \ref{table3} indicates, the results for the same analysis based on a restricted sample (i.e. without the financial crisis) are quite different from the ones presented in table \ref{table2}. In only four out of the 18 countries, a linear growth model with $\widetilde{r}=0$ is chosen. Still, there is a remarkable number of countries with an optimal growth rate which is close to zero and far below commonly discussed growth rates: For four or five countries, (depending on the selection criterion used), the selected growth rate lies in the range between 0 and 1\%.

\begin{table}[t b]
\begin{center}
\begin{tabular}{lccc}
\hline
Country & AIC & AICc & BIC \\ 
\hline
Australia & 0.0245 & 0.0245 & 0.0245 \\ 
Austria & 0.0086 & 0.0086 & 0.0086 \\ 
Belgium & 0.0012 & 0.0012 & 0.0012 \\ 
Canada & 0.0000 & 0.0000 & 0.0000 \\ 
Denmark & 0.0037 & 0.0037 & 0.0037 \\ 
Finland & 0.0282 & 0.0282 & 0.0282 \\ 
France & 0.0000 & 0.0000 & 0.0000 \\ 
Italy & 0.0000 & 0.0000 & 0.0000 \\ 
Luxembourg & 0.0441 & 0.0416 & 0.0416 \\ 
Netherlands & 0.0159 & 0.0159 & 0.0159 \\ 
New Zealand & 0.0441 & 0.0441 & 0.0441 \\ 
Norway & 0.0110 & 0.0110 & 0.0110 \\ 
Portugal & 0.0000 & 0.0000 & 0.0000 \\ 
Spain & 0.0012 & 0.0012 & 0.0012 \\ 
Sweden & 0.0220 & 0.0220 & 0.0220 \\ 
Switzerland & 0.0331 & 0.0331 & 0.0331 \\ 
United Kingdom & 0.0318 & 0.0318 & 0.0318 \\ 
United States & 0.0074 & 0.0074 & 0.0110 \\ 

\end{tabular}
\end{center}
\protect\caption{Optimal growth rates leading to the optimal ARIMA ($p,1,q$) model with drift with respect to different model selection criteria. A growth rate of (close to) zero implies a linear growth model. Time horizon: 1960-2007.}
\label{table3}
\end{table}

Another robustness check investigates only the years 1970-2013. The results in Table \ref{table4} are qualitatively very similar to the ones obtained for the entire sample period. Only three out of 17 countries (excluding Italy which again exhibits ambiguous results) depict exponential growth with a growth rate higher than 1\%.

\begin{table}[t b]
\begin{center}
\begin{tabular}{lccc}
\hline
Country & AIC & AICc & BIC \\ 
\hline
Australia & 0.0172 & 0.0172 & 0.0172 \\ 
Austria & 0.0000 & 0.0000 & 0.0000 \\ 
Belgium & 0.0000 & 0.0000 & 0.0000 \\ 
Canada & 0.0000 & 0.0000 & 0.0000 \\ 
Denmark & 0.0000 & 0.0000 & 0.0000 \\ 
Finland & 0.0000 & 0.0000 & 0.0000 \\ 
France & 0.0000 & 0.0000 & 0.0000 \\ 
Italy & 0.0551 & 0.0000 & 0.0000 \\ 
Luxembourg & 0.0000 & 0.0000 & 0.0000 \\ 
Netherlands & 0.0000 & 0.0000 & 0.0000 \\ 
New Zealand & 0.0135 & 0.0135 & 0.0135 \\ 
Norway & 0.0000 & 0.0000 & 0.0000 \\ 
Portugal & 0.0000 & 0.0000 & 0.0000 \\ 
Spain & 0.0000 & 0.0000 & 0.0000 \\ 
Sweden & 0.0233 & 0.0135 & 0.0135 \\ 
Switzerland & 0.0012 & 0.0012 & 0.0012 \\ 
United Kingdom & 0.0000 & 0.0000 & 0.0000 \\ 
United States & 0.0000 & 0.0000 & 0.0000 \\

\end{tabular}
\end{center}
\protect\caption{Optimal growth rates leading to the optimal ARIMA ($p,1,q$) model with drift with respect to different model selection criteria. A growth rate of (close to) zero implies a linear growth model. Time horizon: 1970-2013.}
\label{table4}
\end{table}

\clearpage

\section{Conclusion}\label{chapfive}
The overall picture of our analyses casts doubts on the widespread belief of exponential growth. The results are not distinct in the sense that we could find clear evidence whether growth in economically developed countries is in general exponential or linear. We observe, however, high dependence on the sample period, in particular on the inclusion of the recent crisis (2008-2013). Besides, our modelling approach is somehow restricted as we only consider linear and exponential trends in ARIMA ($p,1,q$) models. For some of our selected countries other models might reflect the development of their yearly GDPPC  more accurately. With regard to these caveats, we do not want to overreach in the interpretation of our results. Nevertheless, while using more appropriate time series models and more recent data, our main finding is in line with \citet*{Wibe2006} and \citet*{Bourcarde2006}: In contrast to the prominent view of exponential economic growth, a constant growth might be closer to the truth of what has happened in some mature economies within the last 40-50 years .


\bibliographystyle{apa}
\bibliography{Mendeley_References}

\clearpage
\section*{Appendix}\label{chapsix}
 \begin{table}[h t b]
\protect\caption{Unit root tests for the original GDPPC series and its logarithms.}
\label{table_appendix}
\begin{center}
\begin{tabular}{p{4cm}p{2cm}p{2cm}p{2cm}p{2cm}}

 & ADF-Test & ADF-Test & KPSS Test &  KPSS Test \\ 
Country & log. series & orig. series & log. series & orig. series \\
 \hline
\\
Australia & 0.2840 & 0.9024 & 0.0251 & 0.0100 \\ 
Austria & 0.7787 & 0.4320 & 0.0100 & 0.1000 \\ 
Belgium & 0.9221 & 0.9182 & 0.0100 & 0.0126 \\ 
Canada & 0.3972 & 0.3698 & 0.0100 & 0.0265 \\ 
Denmark & 0.8942 & 0.8433 & 0.0100 & 0.0281 \\ 
Finland & 0.6531 & 0.1995 & 0.0100 & 0.0858 \\ 
France & 0.6775 & 0.9689 & 0.0100 & 0.0100 \\ 
Italy & 0.9900 & 0.9900 & 0.0100 & 0.0100 \\ 
Luxembourg & 0.7471 & 0.5411 & 0.0100 & 0.0100 \\ 
Netherlands & 0.3313 & 0.5600 & 0.0100 & 0.0135 \\ 
New Zealand & 0.3428 & 0.4783 & 0.0100 & 0.0100 \\ 
Norway & 0.9900 & 0.8900 & 0.0100 & 0.0180 \\ 
Portugal & 0.6544 & 0.6470 & 0.0100 & 0.0200 \\ 
Spain & 0.9182 & 0.3822 & 0.0100 & 0.0612 \\ 
Sweden & 0.2203 & 0.5959 & 0.0100 & 0.0100 \\ 
Switzerland & 0.0237 & 0.0227 & 0.1000 & 0.1000 \\ 
United Kingdom & 0.5463 & 0.4665 & 0.0452 & 0.0100 \\ 
United States & 0.6881 & 0.5351 & 0.0100 & 0.0100 \\ 
   \hline

\end{tabular}
\end{center}
\caption*{A linear trend in a log-transformed series corresponds to an exponential growth. The KPSS Test in R does not report p-values smaller than 0.01 or bigger than 0.1.}
\end{table}

\end{document}